\documentclass[aps,prl,a4paper,twocolumn]{revtex4}

\usepackage{amsmath,graphicx}

\begin{document}

\title{Quantum state tomography of dissociating molecules}

\author{Esben Skovsen$^1$, Henrik Stapelfeldt$^1$, S\o ren Juhl$^2$ and
Klaus M\o lmer$^3$} \affiliation{1. Department of Chemistry,
University of Aarhus, DK 8000 Aarhus C., Denmark\\ 2. Niels Bohr
Institute, University of Copenhagen, DK 2100 Copenhagen K, Denmark
\\ 3. QUANTOP, Danish
National Research Foundation Center for Quantum Optics,\\
Department of
Physics and Astronomy, University of Aarhus, DK-8000 Aarhus C, Denmark}

\begin{abstract}
Using tomographic reconstruction we determine the complete
internuclear quantum state, represented by the Wigner function, of
a dissociating I$_2$ molecule based on femtosecond time resolved
position and momentum distributions of the atomic fragments. The
experimental data are recorded by timed ionization of the
photofragments with an intense 20 fs laser pulse. Our
reconstruction method, which relies on Jaynes' maximum entropy
principle, will also be applicable to time resolved position or
momentum data obtained with other experimental techniques.
\end{abstract}

\maketitle

Quantum state tomography derives its name from the tomographic
technique in medical diagnostics by which three-dimensional images
of the inner parts of an object or a person can be derived from
two-dimensional NMR or X-ray pictures obtained from different
directions, a technique for which Cormack and Hounsfield were
awarded the Nobel prize in physiology and medicine in 1979. In
quantum physics, we often deal with a phase space,
characterizing the position and velocity distribution of
particles, and the aim of quantum state tomography is to determine
this distribution from only position and momentum measurements on
the particles. As an example, the oscillatory motion of a particle
in a quadratic potential is described as a simple rotation in
phase space, and hence the mathematical Radon-transform technique
\cite{Natterer}, used in physiology and medicine can be used to
extract the phase space distribution of such a particle if only
position measurements are made sufficiently often over a time
corresponding to the oscillation period. This technique has been
succesfully applied to analyze bound molecular states, where the
internuclear distance oscillates harmonically \cite{Whamsley}, and
it has been used for non-classical states of light
\cite{Raymer,Mlynek}, which are formally described as harmonically
trapped particles.  A number of extensions have been made to
incorporate anharmonicities in the trapping potential, and to use
mathematical reconstruction techniques which match specific
detection schemes \cite{Leonhardt,Vogel,Wineland,Haroche}. In the
present work we determine the internuclear quantum state of
dissociating $I_2$ molecules. The atomic fragments
are created in a state with positive energy, and they fly apart
with no confining force between them. The situation thus differs
from the one of trapped motion, and it calls for a new theoretical
approach.

\begin{figure}[here]

 \includegraphics[width = 80 mm]{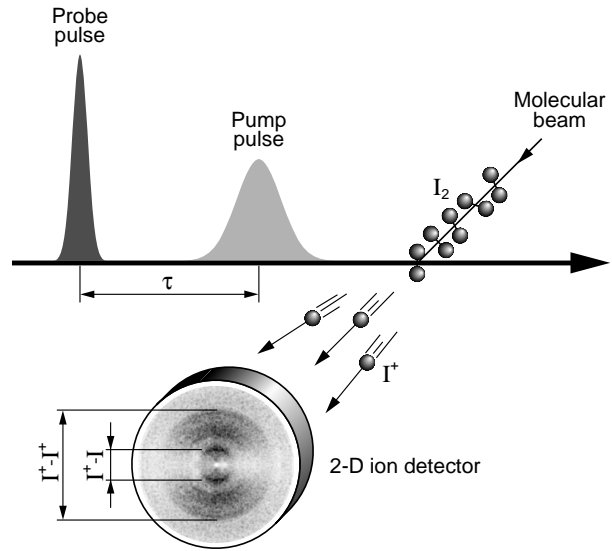}

 \caption{Diagram of the experimental setup showing the molecular beam crossed
  at 90$^o$ by the pump and the probe laser beams. Both laser beams are polarized vertically. The $I^+$ ions produced are
   pushed by a weak electrostatic field (not shown) towards a 2-D ion detector,
   where their positions are used to determine the internuclear
velocity distribution ($I^+$-$I$ channel) and the internuclear
separation distribution ($I^+$-$I^+$ channel). The image shown on
the detector is recorded for a delay of 5 ps between the pump and
the probe pulse. } \label{setup}
 \end{figure}

Experimentally, $I_2$ molecules in a molecular beam are
photodissociated by irradiation with a 100-fs-long pump pulse
centered at 488 nm (see Fig. \ref{setup}). To determine the
internuclear separation (position) and velocitiy (momentum) of the
dissociating molecules at time $\tau$ after excitation the atomic
$I$ fragments are ionized by a 20-fs-long probe pulse, and the
velocities of the $I^+$ ions are measured by a 2-dimensional ion
detector \cite{skovsen}. If only one of the iodine atoms of a
dissociating molecule is ionized the $I^+$ velocity is equal to
half of the internuclear velocity, approximately 4.0 \AA/ps. On
the 2-D image, shown in Fig. \ref{setup}, these ions constitute
the innermost pair of half rings ($I^+$-$I$). If, instead, both
iodine atoms are ionized the velocity of the $I^+$ ions is
increased due to their internal Coulomb repulsion. On the 2-D
image these ions constitute the outermost pair of half rings
($I^+$-$I^+$). Using Coulomb's law the $I^+$ velocity distribution
from this channel gives directly the internuclear distribution
\cite{skovsen}.

The atomic density is sampled on a discrete position grid with $N$
grid points, where $N=51$ in the example shown in Fig. 2.
For the data presented here, the
reconstruction is based on position distributions recorded at
$\tau$ = 2, 3, 4, 5 picoseconds (ps) and the momentum distribution
(which is identical at the four different delays).
In quantum tomography on harmonic oscillator states the fast and
slow velocity components are slushing back and forth, and the time
dependent position distribution maps the phase space distribution
from all angles. In free expansion, slow particles will not be
able to catch up on fast ones, and we have access to only sheared
pictures of the distribution where the high velocity components
are shifted towards larger displacements than the low velocity
components.  This implies that a range of angular viewing angles
is missing, and we cannot apply the Radon transform to reconstruct
the phase space distribution.

Furthermore, the data-sets are noisy due to the finite counts in
different position intervals, and our task is to establish the
most reliable estimate of $\rho$ in best possible
(rather than perfect) agreement with the observed data. Following
Drobny and Buzek \cite{Drobny}, we apply Jaynes' maximum entropy
principle which at the same time provides the smallest deviation
from the measured data and the largest possible von Neumann
entropy \cite{Drobny,Jaynes}: $S(\rho)=-Tr(\rho\ln\rho)$.
Drobny and Buzek have used the maximum entropy principle for quantum
state tomography on harmonically trapped atoms detected at too few
instants of time to allow a normal reconstruction
\cite{Drobnyatom}, and we have reformulated their method to deal
with free atoms.

To calculate the von Neumann entropy, one may
use the familiar expression in terms of the eigenvalues $p_j$
of the density matrix, $S=-\sum_j p_j\ln p_j$, but as it was shown by
Jaynes \cite{Jaynes},
it is not necessary to compute the entropy of $\rho$ in order to use the
maximum entropy principle: The density matrix
with largest entropy which conforms best with measured values $a_i$
for a set of observables $A_i$, can be written on the form
\begin{equation}
\rho = \frac{1}{Z}\exp(-\sum_i \lambda_i A_i),
\label{ansatz}
\end{equation}
where $\lambda_i$ are variables that have to be adjusted
to satisfy the agreement with the measured data, and $Z$
normalizes the density matrix to unit trace.

A position distribution is the set of expectation values of
projection operators on position eigenstates $|x\rangle\langle x|$.
We determine the position distribution at different times,
and we need a formal representation of the projection
operator $|x,t\rangle\langle x,t|$ where $|x,t\rangle$ is a position
eigenstate at time $t$.  For free particles the  momentum $p=\hbar k$ is a
conserved quantity, and it is convenient to represent the projection
operators in a momentum representation,
since momentum eigenstates accumulate only a trivial phase factor with time,
$|k,t\rangle = \exp(i\hbar k^2t/2M)|k,t=0\rangle$:
\begin{eqnarray}
|x\rangle \langle x|=\int dk_1 dk_2 |k_1\rangle\langle
k_1|x\rangle \langle x|k_2\rangle\langle k_2|\nonumber \\
=
\frac{1}{2\pi}\int dk_1 dk_2 \exp(i(k_2-k_1)x) |k_1\rangle\langle k_2|.
\end{eqnarray}
Both the spatial distribution at various times and the momentum
distribution, which is independent of time, are measured, and the
observables $A_i$ in Eq.(1) are precisely the  operators
$|x,t\rangle\langle x,t|$ and $|k\rangle\langle k|$, which are now
given explicity as $N$ by $N$ matrices in a discrete basis. The
exponential in Eq.(1) is understood as a matrix power series, and
it has to be evaluated numerically because the momentum projection
operators and the position projection operators at different times
do not commute.

In our numerical reconstruction we used four position
distributions and one momentum distributions, and we thus
identified $5\cdot 51$ variables $\lambda_i$ so that the
expectation values $\langle A_i\rangle=$Tr$(\rho A_i)$ are as
close as possible to the measured distributions. We quantify the
agreement with the measured data $a_i$ by the sum of the squares
of all deviations $\Delta = \sum_i(a_i-\langle A_i\rangle)^2$. A
numerical routine identifies the global minimum of $\Delta$ as a
function of the $\lambda_i$ variables, and we thus obtain the
explicit expression for the density matrix  in Eq.(1).

To illustrate our reconstruction of the internuclear quantum state
of the dissociated molecules we use the Wigner function  $W(x,p)$
\cite{Hillery} rather
than the N by N complex density matrix, discussed above.
The Wigner function  is obtained by a Fourier transform with respect
to the difference between the two momentum
arguments in the density matrix. While
still fully characterizing the quantum state the Wigner function
is a real function, and it has the advantage of presenting at one
glance the joint position and momentum distribution. Figure 2
shows the Wigner function, $W(x,p)$, at the earliest (2 ps) and
the latest time (5 ps) for which position distributions were
recorded. Note that we present the momentum dependence  in units
of the velocity $v=p/M$ of the atomic fragments to facilitate
comparison of the two panels. The two Wigner functions in Fig. 2
a) and b) have the same velocity distributions, but they are
sheared differently in phase space due to the free motion of the
two iodine photofragments.  On the side panels the marginal
position and velocity  distribution, obtained by integrating
$W(x,p)$ over velocity and position, respectively, are compared
with the experimental results (open circles). The agreement is
excellent, and similar good agreement is found between the
measured data and the reconstructed state at the intermediate
times at 3 and 4 ps.

 \begin{figure}[htbp]

 \includegraphics[width = 80 mm]{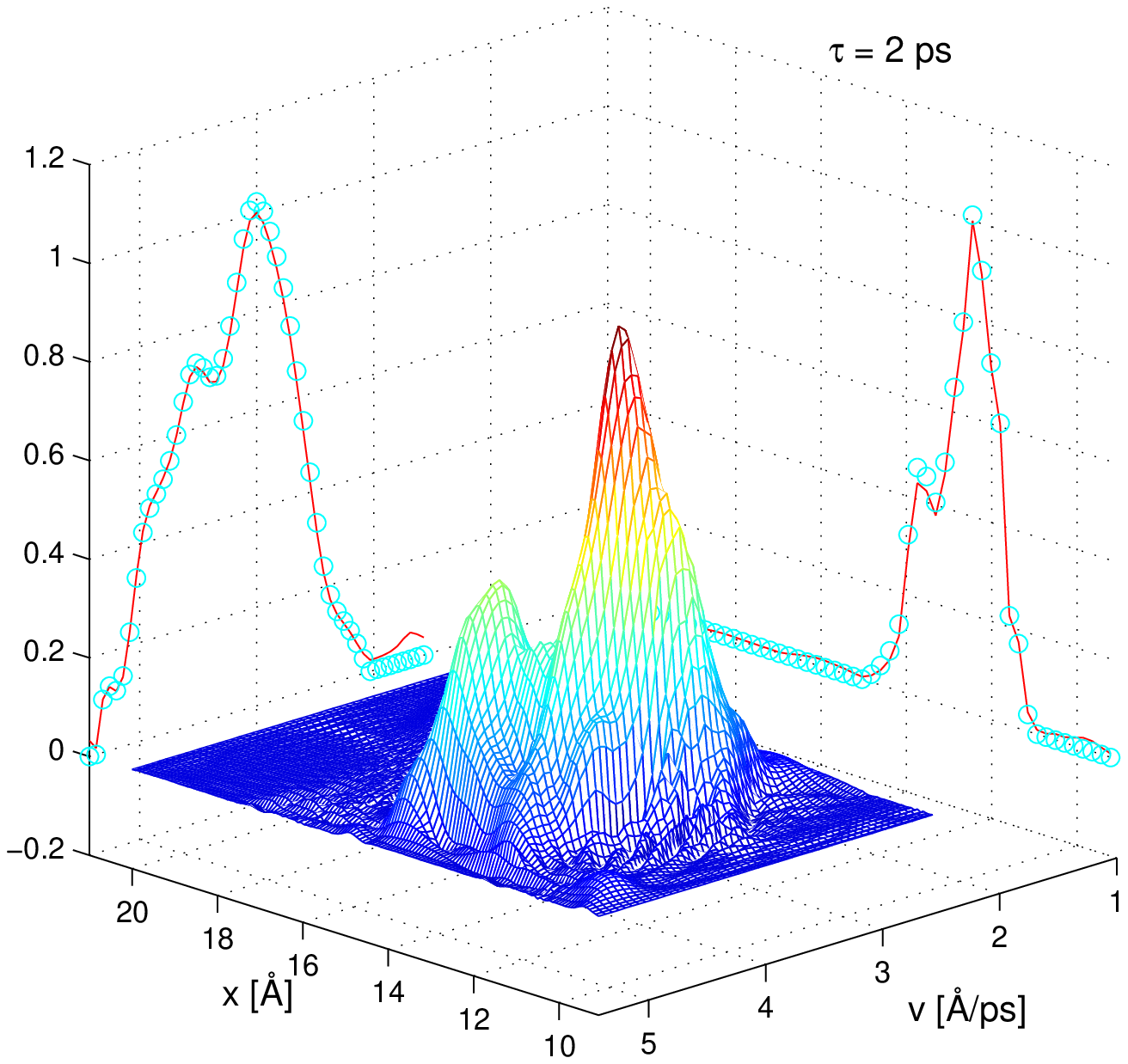}
\includegraphics[width = 80 mm]{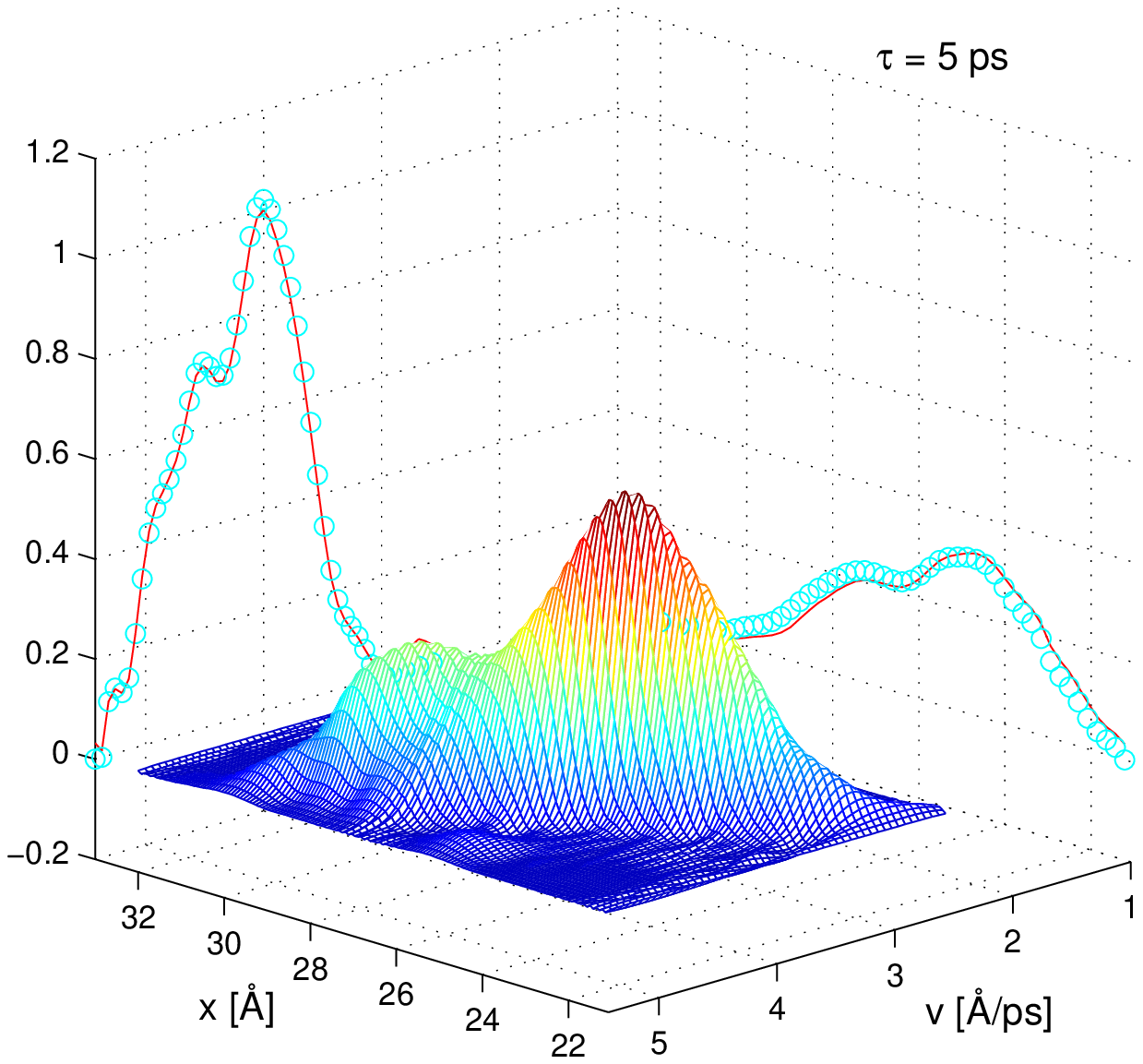}
 \caption{Three dimensional surface plots showing the phase space
distribution function $W(x,p)$ (Wigner function) at two different
times after the pump pulse dissociates the $I_2$ molecule (tau=2ps
in part (A) and tau = 5 ps in part (B)). The marginal position and
momentum distributions of the dissociated molecules are shown on
the side panels: The full curves are the reconstructed
distributions (integrals over the Wigner function with respect to
momentum and position, respectively), and the open circles are the
measured distributions. To provide a more intuitive view the
momentum variable is presented in velocity units. }
 \label{wigner}
 \end{figure}

We stress that the Wigner function contains more information than
what can be obtained directly from a single measured position and
velocity distribution. For instance, both the measured position
and velocity distribution at 2 ps consists of two peaks (side
panels of Fig. 2 (A)), but without further information it is not
possible to determine if the molecules at the larger internuclear
separation (the peak around 13.6 \AA\ ) are also the ones moving
with the largest internuclear velocity (the peak around 4.6
\AA\//ps). The tomographic reconstruction of the Wigner function
makes use of our knowledge of the position distributions measured
at 3, 4 and 5 ps, and, indeed, the two separated peaks in $W(x,p)$
at 2 ps show that the iodine atoms roughly 13.6 \AA\ apart have a
velocity distribution centered around 4.6 \AA\//ps whereas the
iodine atoms apparoximately 12.4 \AA\ apart have a velocity
distribution centered at 4.0 \AA\//ps. The physical origin of the
two internuclear velocity components is the incoherent population
of v = 0 and v = 1 vibrational levels of the electronic ground
state of the I$_2$ molecules prior to dissociation by the pump
pulse \cite{skovsen}.

It has been emphasized in the literature \cite{Hillery} that the
Wigner function is not a real probability distribution because, due to
complimentarity and Heisenbergs's uncertainty relation, it is not
meaningful to assign precise values for both the position and
velocity of a particle. This has as its most striking consequence
that states exist for which $W(x,p)$ attains negative values, but
these negative values always occur in phase space regions with
area less than Planck's constant $\hbar=h/2\pi$. Dissociation of
molecules by two phase-locked femtosecond laser pulses will
produce photofragments in a coherent superposition of two
localized wave packets \cite{skovsen}, and we are currently
working on improvement of the experiments to enable reconstruction
of the Wigner function with a sufficient resolution to display the
negativities expected in this case. Such an experiment constitutes
a fs time resolved analogue to double-slit atom interferometer
studies \cite{Mlyneknature}.

Quantum state tomography has been established in quantum optics as
a remarkable diagnostics tool that has provided clear illustrations
of the experimental ability to control and manipulate selected
states of fundamental quantum systems such as a single field mode
and single ions and atoms \cite{Mlynek, Vogel, Drobnyatom, Wineland, Haroche}.
Our experimentally reconstructed density matrix or
Wigner function provides the complete information about the
non-trivial quantum state of the internuclear motion of the
dissociating $I_2$ molecules. Although our experimental method is
limited to small molecules, we believe that quantum state
tomography will be applicable also to study chemical reactions in
large molecules through time resolved position data obtained,
e.g., by the emerging femtosecond electron and X-ray diffraction
techniques \cite{Ihee, Reutze, Corkum, Service}. This will open
for comparisons with theoretical calculations at a much more
detailed level than probability distributions of single
observables, which have formed the basis of comparison so far in
femtosecond time resolved chemical reaction dynamics. Finally, we
note that modern femtosecond laser technology enables controlled
shaping of both the electronic and the atomic structure of
molecules through irradiation with sequences of tailored laser
pulses \cite{Rabitz, Levis, Gerber}. To fully exploit the
capabilities of such quantum manipulation it is crucial to
completely characterize the molecular quantum states formed.

We acknowledge the support from the Carlsberg Foundation and The 
Danish Natural Science Council.

\vspace*{100mm}
\vfill
\eject

\end{document}